\documentclass[letterpaper, 10 pt, conference]{ieeeconf}  % Comment this line out if you need a4paper

\IEEEoverridecommandlockouts                              % This command is only needed if 
\usepackage[T1]{fontenc}
\usepackage{graphics} % for pdf, bitmapped graphics files
\usepackage{epsfig} % for postscript graphics files
\usepackage{mathptmx} % assumes new font selection scheme installed
\usepackage{times} % assumes new font selection scheme installed
\usepackage{amsmath} % assumes amsmath package installed
\usepackage{amssymb}  % assumes amsmath package installed
\usepackage{tabularx}
\usepackage{graphicx}
\usepackage{multirow}
\usepackage{arydshln} 
\usepackage[symbol]{footmisc}

\title{\LARGE \bf
Extending Stress Detection Reproducibility to \\ Consumer Wearable Sensors
}
\author{Ohida Binte Amin\textsuperscript{1, 2}, Varun Mishra\textsuperscript{1, 2}, Tinashe M. Tapera\textsuperscript{1, 2}, Robert Volpe\textsuperscript{2}, Aarti Sathyanarayana\textsuperscript{1, 2, 3} \\
\textsuperscript{1}Khoury College of Computer Sciences, Northeastern University, Boston, MA\\
\textsuperscript{2}Bouvé College of Health Sciences, Northeastern University, Boston, MA\\
\textsuperscript{3}Department of Biostatistics, Harvard T.H. Chan School of Public Health, Boston, MA\\
{\tt\small \{amin.o, v.mishra, tapera.t, r.volpe, a.sathyanarayana\}@northeastern.edu}
}
\begin{document}

\maketitle
\thispagestyle{empty}
\pagestyle{empty}

%%%%%%%%%%%%%%%%%%%%%%%%%%%%%%%%%%%%%%%%%%%%%%%%%%%%%%%%%%%%%%%%%%%%
\begin{abstract}
Wearable sensors are widely used to collect physiological data and develop stress detection models. However, most studies focus on a single dataset, rarely evaluating model reproducibility across devices, populations, or study conditions. We previously assessed the reproducibility of stress detection models across multiple studies, testing models trained on one dataset against others using heart rate (with R-R interval) and electrodermal activity (EDA). In this study, we extended our stress detection reproducibility to consumer wearable sensors. We compared validated research-grade devices, to consumer wearables - Biopac MP160, Polar H10, Empatica E4, to the Garmin Forerunner 55s, assessing device-specific stress detection performance by conducting a new stress study focused on solely undergraduate students at Northeastern University. Thirty-five students completed three standardized stress-induction tasks in a controlled lab setting. Biopac MP160 performed the best, being consistent with our expectations of it as the gold standard, though performance varied across devices and models. Combining heart rate variability (HRV) and EDA enhanced stress prediction across most scenarios. However, Empatica E4 showed variability; while HRV and EDA improved stress detection in leave-one-subject-out (LOSO) evaluations (AUROC up to 0.953), device-specific limitations led to underperformance when tested with our pre-trained stress detection tool (AUROC 0.723), highlighting generalizability challenges related to hardware-model compatibility. Garmin Forerunner 55s demonstrated strong potential for real-world stress monitoring, achieving the best mental arithmetic stress detection performance in LOSO (AUROC up to 0.961) comparable to research-grade devices like Polar H10 (AUROC 0.954), and Empatica E4 (AUROC 0.905 with HRV-only model and AUROC 0.953 with HRV+EDA model), with the added advantage of consumer-friendly wearability for free-living contexts. Exploring potential devices that work effectively in a controlled lab setting could significantly enhance the reliability of stress detection tools in real-world applications. \footnote[1]{Accepted at IEEE EMBC 2025}
\newline

\end{abstract}

%%%%%%%%%%%%%%%%%%%%%%%%%%%%%%%%%%%%%%%%%%%%%%%%%%%%%%%%%%%%%%%%%%%%%%%%%%%%%%%%

\section{INTRODUCTION}
Effective stress monitoring and management are essential for maintaining overall well-being, especially for students who frequently encounter high stress due to persistent academic demands and social expectations \cite{vidal2022fluctuations}. Proactively managing stress reduces the risk of stress-related health issues like cardiovascular disease, gastrointestinal disorders, substance abuse, and even chronic conditions including diabetes, and hypertension, enhancing quality of life \cite{can2019stress}. The increasing availability of wearable devices has facilitated continuous stress monitoring, enabling studies with minimal participant burden \cite{vidal2022fluctuations}. Physiological biomarkers such as heart rate variability (HRV) and electrodermal activity (EDA), measured through digital devices, provide reliable indicators of stress responses governed by the Autonomic Nervous System and the Hypothalamic-Pituitary-Adrenal axis \cite{vos2023generalizable}.

Stress assessment has evolved through both controlled lab experiments and real-world, free-living conditions. In controlled lab settings, leveraging a variety of well-validated tasks (such as mental arithmetic, the Stroop test, startle response tests, cold-pressor tests, or public speaking), prior studies were designed to elicit physiological stress responses under different stress conditions while incorporating restful periods as a baseline \cite{mishra2020continuous, king2019micro, egilmez2017ustress, gjoreski2017monitoring, hovsepian2015cstress, sarker2016finding, chen2019temporal, plarre2011continuous}. In contrast, stress levels were monitored in uncontrolled free-living situations where several contactless methods were employed for stress measurement, including analyzing user's voice \cite{lu2012stresssense}, keyboard typing behaviors \cite{saugbacs2020stress} or using accelerometer-based contextual modeling \cite{garcia2015automatic}, or smartphone data like Bluetooth, app usage, and Call/SMS logs \cite{bogomolov2014pervasive, sano2013stress, garcia2015automatic, wang2014studentlife}. Many studies used custom-built sensing systems for precise control over signal quality, sensor type, and battery life \cite{choi2011development, greco2015cvxeda, hovsepian2015cstress, plarre2011continuous, sun2012activity}, but these limit scalability and broader validation. Alternatively, some studies leveraged commercially available sensors, either alone or combined with custom sensors or smartphones \cite{sano2013stress}. Wearable technology has the potential to consistently obtain high-quality biosignals over weeks to months and is increasingly valued for its advanced diagnostic and therapeutic potential \cite{stuart2022wearable}. Muaremi et al. \cite{muaremi2014monitoring} monitored sleep stress using both the Zephyr BioHarness 3.0 and an Empatica E3 \cite{garbarino2014empatica}. Gjoreski et al. \cite{gjoreski2017monitoring} detected stress in various environments by employing both the Empatica E3 and E4 \cite{empatica-e4}. Egilmez et al. \cite{egilmez2017ustress} \& Mishra et al. \cite{mishra2020continuous} used a commodity device the Polar H7 heart rate monitor for continuous detection of physiological stress. Pinge et al. compared the performance of a Garmin Vivosmart 4 Smartwatch \cite{Garmin_Vivosmart}, a Polar H10 \cite{Polar_10}, and a Garmin HRM Dual \cite{Garmin_hrm} Heart Rate Monitor for stress detection. 

Advanced machine learning and deep learning techniques have enhanced the effectiveness of stress detection using data from wearable devices. Regarding advancement in stress detection models, Hovsepian et al. used data from a lab study with 21 participants to build a stress-detection model, cStress using a Support Vector Machine (SVM) model with a Radial Basis Function (RBF) kernel \cite{hovsepian2015cstress}, while Sarker et al. applied the same cStress model to identify stress among 38 polydrug users \cite{sarker2016finding}. In another study, Heyat et al. developed an automatic stress detection system using Electrocardiogram (ECG) signals from smart T-shirts with machine learning classifiers, Decision Tree (DT), Naive Bayes (NB), Random Forest (RF), Logistic Regression (LR) including data from 20 subjects \cite{bin2022wearable}. Vito et al. designed a real-time stress detection system applying LR along with Ridge LR, DT, RF, and SVM with RBF kernel using only Heart Beats Per Minute \cite{di2023development}. Bari et al. introduced a stress detection model using LR and RF for conversations incorporating ECG and respiration sensors to gather physiological stress data, encompassing both controlled lab and real-world environments \cite{bari2020automated}. Researchers aimed to design a real-time method to identify stress episodes from time series data of physiological stress markers, enabling the prompt initiation of just-in-time stress-management interventions. 

While these works have contributed to stress detection advancements, significant challenges remain, particularly regarding reproducibility that might lead to the generalizability of stress detection models. Here, \textbf{reproducibility} refers to a scenario in which a team of researchers applies models or methods from a previous study and achieves similar results in an independent study with a different experimental setup \cite{mishra2020evaluating}. Stress is inherently subjective, and its physiological manifestations might vary across individuals. Different people may exhibit distinct autonomic and hormonal responses to the same stressor, and their self-reported experiences of stress may not always align with physiological measurements. This variability complicates the development of generalizable stress-detection tools. Despite advancements in wearable-based stress detection, many studies develop stress detection models using physiological data from controlled settings, often with custom or commercial wearables. However, these models are typically trained and validated within a single study, limiting their reproducibility to new datasets, populations, or devices. A major limitation is the lack of systematic efforts to assess whether models trained on one dataset remain effective across different studies, participant groups, or sensor types. Inconsistencies in data collection methods, sensor placement, and preprocessing further complicate cross-study comparisons. 

Our prior work explored the reproducibility of stress detection models by evaluating their performance across multiple studies and two different sensor types, Polar H7/H10 chest monitor and Empatica E4 devices \cite{mishra2020evaluating}. We evaluated the performance of models built using data from one study and tested on data from other studies.  In this study, to evaluate consumer wearable device-specific reproducibility, we conducted a digital phenotyping stress study with 35 undergraduate students at Northeastern University. We assessed the feasibility of different wearable devices for stress detection using our previously developed stress detection tool across multiple devices—Polar H10, Empatica E4, Garmin Forerunner 55s, and Biopac MP160 (as a gold-standard medical device). We specifically focused on students, as they frequently experience academic and social stress in daily life, to extend stress detection reproducibility to the consumer wearables Garmin Forerunner 55s, given its increasing acceptance as a health-tracking device \cite{evenson2020review}. We assessed the performance of devices within the study in capturing physiological stress markers using leave-one-subject-out (LOSO) cross-validation and further tested the stress versus restful periods from our 35 students with our pre-trained stress model \cite{mishra2020evaluating}. This comparative analysis offers insights into the potential of commercially available wearables, particularly the Garmin Forerunner 55s, to enhance the reproducibility of our stress detection tool tailored for student populations.

\begin{table}[h]
\centering
\caption{Socio-demographics of 35 undergraduate students}
\label{tab:participant_demographics}
\begin{tabular}{lcc}
\hline
\textbf{Characteristic} & \textbf{Number of Participants} & \textbf{Percentage (\%)} \\
\hline
\multicolumn{3}{l}{\textit{Age Group}} \\
Below 20 & 7 & 19.4\% \\
20 to 22 & 22 & 63.9\% \\
Above 22 & 6 & 16.7\% \\
\hline
\multicolumn{3}{l}{\textit{Gender Identity}} \\
Female & 21 & 58.3\% \\
Male & 13 & 38.9\% \\
Non-Binary & 1 & 2.8\% \\
\hline
\multicolumn{3}{l}{\textit{Hispanic/Latino Identity}} \\
Yes & 6 & 16.7\% \\
No & 29 & 83.3\% \\
\hline
\multicolumn{3}{l}{\textit{Academic Year}} \\
Freshman & 4 & 11.1\% \\
Sophomore & 14 & 41.7\% \\
Junior & 8 & 22.2\% \\
Senior & 9 & 25\% \\
\hline
\end{tabular}
\end{table}

\section{Methodology}
While stress detection using wearables has advanced, models often lack reproducibility across studies, populations, and devices due to variations in data collection, sensor placement, and physiological data preprocessing. 
To address this, we previously evaluated the reproducibility of stress detection models by testing their performance across multiple studies \cite{mishra2020evaluating}, assessing whether models trained on one dataset could achieve comparable results on another with similar stressors and sensor modalities. However, these challenges continue to hinder the broader applicability of stress monitoring frameworks in real-world settings. Addressing them is crucial for advancing digital phenotyping in stress monitoring and developing reliable, scalable stress management interventions. In the following section, we first briefly discuss how we built our stress detection tool evaluating reproducibility, and later we discuss how we extended our stress detection reproducibility across diverse wearable sensors focusing primarily on consumer wearables, Garmin Forerunner 55s.

\subsection{Approach \& Development of Stress Prediction Tool for Evaluating Reproducibility}

\subsubsection{Physiological Signal Selection \& Pre-processing}
To evaluate the reproducibility of stress detection models, we used heart rate (HR) and R-R interval data from Polar H7/H10 chest monitors and EDA data from the wrist-worn Empatica E4s across four studies. For consistency, only processed HR and HRV data were used from all devices. Physiological data were preprocessed to remove artifacts and standardize across datasets. HR values outside 30–220 bpm and EDA values below 0.01 $\mu$S or above 100 $\mu$S were discarded \cite{mishra2018case}. Motion artifacts from sudden movements were handled separately: HR and HRV outliers were trimmed using a Median Absolute Deviation (MAD) threshold (values beyond median ± 3×MAD were removed), while EDA artifacts were smoothed using a 5-second median filter. To address inter-participant variability, HR and R-R interval data were Z-score normalized, as recommended \cite{mishra2018case}, while EDA signals were min-max normalized to [0,1]. Additionally, EDA signals were decomposed into tonic and phasic components using the cvxEDA method, separating sustained physiological arousal from rapid fluctuations.

\subsubsection{Feature Extraction}
To extract stress-related features, we used a 60-second sliding window with a 45-second overlap. Time-domain features from HR and R-R intervals like mean, standard deviation, skewness, and percentiles were extracted while frequency-domain features were excluded due to their limited benefit in stress classification. For EDA, we computed statistical features from the overall signal and its tonic and phasic components, including peak count, area under the curve (AUC), etc.

\subsubsection{Machine Learning Models and Evaluation}
SVM and RF models are known for their ability to perform better to detect stress in prior literature \cite{egilmez2017ustress, gjoreski2017monitoring, hovsepian2015cstress, king2019micro} by limiting overfitting and reducing the bias and variance tradeoffs. SVM also consistently performed better when using only heart rate and R-R interval features, while RF showed superior performance when incorporating additional features such as EDA \cite{mishra2018case}. 
SVM with RBF kernel (C = 107, \( \gamma = 0.001 \)) and RF with 100 trees were used for LOSO cross-validation during cross-study evaluation. To assess reproducibility, models trained on one dataset were tested on another without hyperparameter tuning. We found that models trained on Polar H7/H10 generalized better than those trained on Empatica E4, suggesting sensor variability affects model transferability.

\subsection{Extending Stress Detection Reproducibility with Multi-Device Testing}

Our prior work on evaluating the reproducibility of stress detection models focused on only two device types (Polar H7/H10 and Empatica E4) \cite{mishra2020evaluating}. This restricts device-specific reproducibility, as physiological stress responses might vary across devices in signal quality, preprocessing, and sensor placement that impact model performance. Expanding our analysis with diverse devices including consumer wearables and narrowing students would enhance the robustness of our stress detection tool. 

We conducted a stress study with undergraduate students recruited from Northeastern University, with eligibility criteria requiring them to be at least 18 years old, fluent in English, and own an iOS 12+ smartphone. Recruitment and study protocols were approved by Northeastern University’s Institutional Review Board (IRB\#22-10-02), and participants were duly compensated at the end of the study. We tested four different sensors: Empatica E4, Garmin Forerunner 55s, Biopac MP160, and Polar H10 chest monitor.

\begin{figure}[h] 
    \centering
    \includegraphics[width=\linewidth]{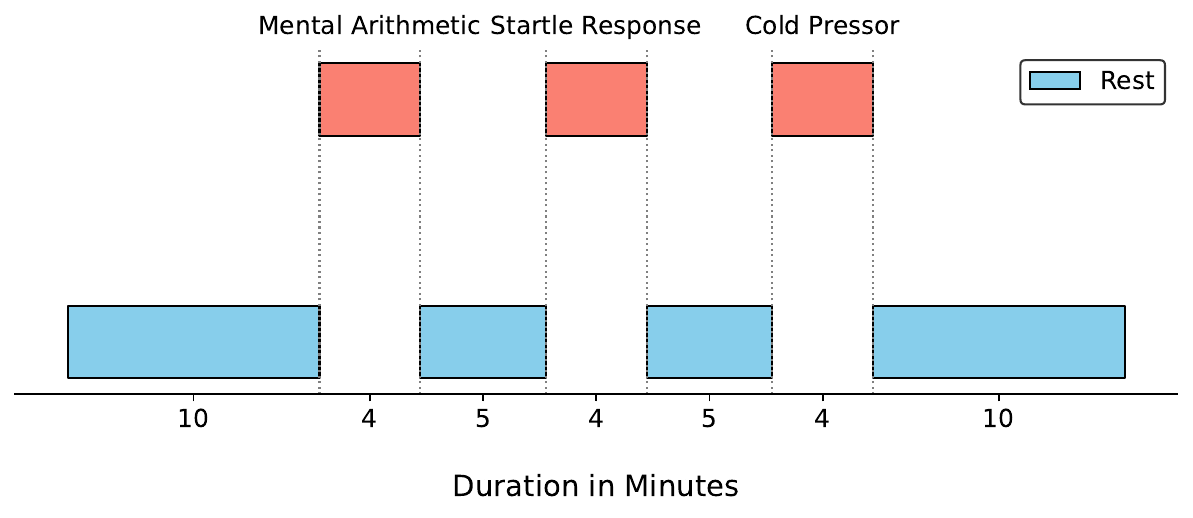} 
    \caption{Lab protocol timeline for three stressor conditions in our digital phenotyping stress study, where the first two stress tasks were randomized. The cold pressor was conducted at the end due to logistical constraints.}
    \label{fig:timeline}
\end{figure}

\subsubsection{In-Lab Calibration and Stress Induction Protocol}
During controlled in-lab sessions, after providing informed consent, participants were fitted with three wearable devices: the Empatica E4, Garmin Forerunner 55, and Polar H10 heart rate monitor. Additionally, ECG and EDA electrodes were attached and connected to the Biopac MP160 system for high-precision physiological data collection.
As illustrated in Fig. \ref{fig:timeline}, the session began with a 10-minute baseline rest period, followed by three stress-inducing tasks, each lasting four minutes and separated by a five-minute rest period. The stress tasks included: (1) a mental arithmetic challenge, where participants counted backward in steps of seven from a large number; (2) a startle response task, involving unexpected loud noises; and (3) a cold pressor task, requiring participants to submerge their hands in ice water.

\subsection{Comparative Analysis of Stress Detection Performance Across Devices}
\begin{table*}[ht]
\caption{Comparative Analysis of Stress Detection Models Across Multiple Devices (Biopac MP160, Polar H10, Empatica E4, and Garmin Forerunner 55s) in Lab Session}
\label{table:results}
\centering
\renewcommand{\arraystretch}{1.5}
\newcolumntype{Z}{>{\centering\arraybackslash}m{3.5cm}}
\begin{tabularx}{\textwidth}{|X|X|X|X|X|}
\hline
\multirow{2}{*}{\textbf{Training Data}} & \multicolumn{2}{c|}{\textbf{LOSO Cross-Validation}} & \multicolumn{2}{c|}{\textbf{Pre-trained Mishra~et~al. Model \cite{mishra2020evaluating}}} \\ \cline{2-5} 
                                         & \textbf{Rest vs. All Stressors} & \textbf{Rest vs. Mental Arithmetic Stressor} & \textbf{Rest vs. All Stressors} & \textbf{Rest vs. Mental Arithmetic Stressor} \\ \hline

Biopac MP160          & \textbf{HRV:} 0.776 [0.566-0.834]\par\noindent\rule{\linewidth}{0.4pt}\par
\textbf{HRV+EDA:} 0.984 [0.908 - 0.995] & \textbf{HRV:} 0.755 [0.50-0.918]\par\noindent\rule{\linewidth}{0.4pt}\par
\textbf{HRV+EDA:} 1.0 [0.941 - 1.0] & \textbf{HRV:} 0.762 [0.523-0.867]\par\noindent\rule{\linewidth}{0.4pt}\par
\textbf{HRV+EDA:} 0.970 [0.926, 0.994] & \textbf{HRV:} 0.794 [0.314-0.926]\par\noindent\rule{\linewidth}{0.4pt}\par
\textbf{HRV+EDA:} 0.999 [0.929, 1.000] \\ \hline
Polar H10             & \textbf{HRV:} 0.73 [0.62-0.84]\newline  &  \textbf{HRV:} 0.954 [0.85-0.983]\newline  &  \textbf{HRV:} 0.729 [0.599-0.844]\newline  & \textbf{HRV:} 0.961 [0.858-0.998]\newline  \\ \hline

Empatica E4           & \textbf{HRV:} 0.61 [0.51-0.77]\par\noindent\rule{\linewidth}{0.4pt}\par
\textbf{HRV+EDA:} 0.884 [0.698-0.944] & \textbf{HRV:} 0.905 [0.848-0.974]\par\noindent\rule{\linewidth}{0.4pt}\par
\textbf{HRV+EDA:} 0.953 [0.606-0.999] & \textbf{HRV:} 0.548 [0.443-0.834]\par\noindent\rule{\linewidth}{0.4pt}\par
\textbf{HRV+EDA:} 0.736 [0.550, 0.865] & \textbf{HRV:} 0.913 [0.835-0.969]\par\noindent\rule{\linewidth}{0.4pt}\par
\textbf{HRV+EDA:} 0.723 [0.401, 0.964] \\ \hline
Garmin Forerunner 55s    & \textbf{HRV:} \textbf{0.73} [0.51-0.81]\newline  & \textbf{HRV:} \textbf{0.961} [0.899-0.998]\vfill  & \textbf{HRV:} \textbf{0.748} [0.509-0.842]\newline  & \textbf{HRV:} \textbf{0.968} [0.886-0.998]\newline  \\ \hline
\multicolumn{5}{l}{The cells denote the median AUROC score across all students, along with the inter-quartile range.}
\end{tabularx}
\end{table*}

Biopac MP160 provided EDA and ECG data during the lab session, with R-R intervals calculated using AcqKnowledge software. Following our prior work on evaluating the reproducibility of stress detection models \cite{mishra2020evaluating}, HR (and R-R interval) data were obtained from all devices, while EDA data were specifically obtained from Empatica E4 and Biopac MP160. We pre-processed these physiological data and extracted useful features. Using our 35 students' in-lab physiological data, we tested the reproducibility of our stress detection tool, which was trained on the prior dataset, and output a probability of stress based on the features from a 60-second window. To enrich our analysis, we also compared the performance of our prior stress detection models with LOSO cross-validation using physiological data collected from multiple devices (Empatica E4, Garmin Forerunner 55s, Biopac MP160, and Polar H10 chest monitor). We had two models as mentioned one was an SVM model that outputs a probability of stress using just heart rate and R-R interval features; the second was an RF model, that outputs an input sample by computing the mean predicted class probabilities of the trees in the forest \cite{mishra2020evaluating}. 
We employed the SVM model for analyses focusing solely on HRV data, irrespective of the device. Conversely, when incorporating both HRV and EDA data, we applied RF to identify stress in controlled lab settings. We assessed the performance of these models using the Area Under the Receiver Operator Characteristics (AUROC) scores, a metric that measures a model's ability to differentiate between two classes independent of changes in class distribution within the test set. 

In our analysis, we considered the following scenarios: \textbf{Scenario 1:} 10 minutes of the first baseline is considered 0, and all three stressors are considered 1. \textbf{Scenario 2}: 10 minutes of the first baseline is considered 0, but only the mental arithmetic is considered 1. We decided to break our evaluations across the two scenarios to enable a fair comparison with the model we developed in prior work, which was trained only using the baseline and mental arithmetic stressors. For our LOSO evaluations, we employed SVM (using the LibSVM library) with RBF kernels and RF algorithm (using the Scikit-learn library) as the primary machine-learning models due to their robust performance in stress detection following the previously mentioned architecture. 

\section{Results}
\subsection{Socio-demographics of Participants in Digital Phenotyping Stress Study}
Table \ref{tab:participant_demographics} presents the socio-demographic breakdown of 35 undergraduate students who participated in our study. The majority of participants are in the age group of 20 to 22 years, making up 63.9\% of the sample. Females are the predominant gender identity represented, accounting for 58.3\%, while males and non-binary individuals comprise 38.9\% and 2.8\%, respectively. A small portion of the sample identifies as Hispanic/Latino, at 16.7\%. In terms of academic standing, sophomores form the largest group (41.7\%), followed closely by seniors (25\%), juniors (22.2\%), and freshmen (11.1\%). This demographic spread suggests a diverse sample across genders and academic years, with a lower representation of ethnic diversity and age variation.

\subsection{Comparative Stress Prediction Reproducibility Performance Across Devices}
Our findings are represented in Table \ref{table:results}, which highlights the predictive capabilities of HRV and EDA in stress detection models. We present the median AUROC scores (along with the Interquartile Range) across all participants. 

In \textbf{Scenario 1} (Rest vs. All Stressors), under LOSO cross-validation, \textbf{Biopac MP160} achieved an AUROC of 0.776 with HRV-only stress detection model while adding EDA significantly improved performance to an AUROC of 0.984, reinforcing the advantage of multimodal sensing. When testing the pre-trained Mishra et al. model with our study's data, Biopac MP160 exhibited slightly lower performance in HRV-only model with an AUROC of 0.762 but maintained a high performance AUROC of 0.970 when incorporating EDA, confirming the robustness of its physiological measurements. In \textbf{Scenario 2} (Rest vs. Mental Arithmetic Stressor), LOSO results showed that Biopac MP160 achieved an AUROC of 0.755  with HRV-only stress detection model, while HRV \& EDA reached a perfect AUROC of 1.0, confirming its superior accuracy. The pre-trained Mishra et al. model yielded similar trends, with an AUROC of 0.794 for HRV-only stress detection model and an AUROC of 0.999 with HRV \& EDA, indicating strong model generalizability for Biopac MP160. However, its cumbersome setup, and lack of portability limit its usability for real-world continuous stress monitoring.

\textbf{In Scenario 1}, LOSO evaluation showed that \textbf{Polar H10} chest monitor achieved an AUROC of 0.73 with its HRV-only stress detection model, demonstrating its effectiveness as a research-grade stress monitoring device. The pre-trained Mishra et al. model achieved a nearly identical performance with an AUROC of 0.729, reinforcing its reliability in stress detection. In \textbf{Scenario 2}, LOSO results demonstrated improved performance for HRV-only model, with Polar H10 achieving an AUROC of 0.954, confirming its ability to effectively capture stressful episodes. Similarly, the pre-trained Mishra et al. model achieved an AUROC of 0.961, as it was originally developed using the mental arithmetic task, further validating the consistency of our prior stress detection tool's reproducibility. However, Polar H10’s chest-worn design may reduce long-term usability, as its fixed wearability position hampers flexibility, making it less practical for daily continuous stress monitoring.

For \textbf{Empatica E4}, in \textbf{Scenario 1}, LOSO evaluation showed lower HRV-only performance, an AUROC of 0.61, but performance significantly improved when combining HRV \& EDA with an AUROC of 0.884, reinforcing the importance of multimodal sensing, as seen with Biopac MP160. However, the pre-trained Mishra et al. model exhibited poor generalization, with HRV-only model achieving an AUROC of 0.548, which, for a binary classification task, is almost equivalent to random guessing. While adding HRV \& EDA improved performance with an AUROC of 0.736, it remained lower than other HRV \& EDA models, indicating the model's lack of generalizability with Empatica E4’s data. In \textbf{Scenario 2}, LOSO results demonstrated strong HRV-only performance with an AUROC of 0.905 and HRV \& EDA reaching an AUROC of 0.953, comparable to Polar H10. However, the pre-trained Mishra et al. model again failed to generalize well, with HRV \& EDA performance dropping to AUROC of 0.723 compared to its' HRV-only model AUROC of 0.913, further questioning the reliability of Empatica E4’s EDA sensor for real-world stress tracking. Moreover, Empatica E4 is not a proper consumer wearable as it is primarily used for research purposes and lacks widespread consumer adoption.

Lastly, for \textbf{Garmin Forerunner 55s}, in \textbf{Scenario 1}, LOSO evaluation showed strong HRV-only stress detection performance with an AUROC of 0.73, demonstrating its feasibility for stress detection, comparable to Polar H10. The pre-trained Mishra et al. model obtained similar results, with an AUROC of 0.748 for its HRV-only model, reinforcing its potential in real-world applications. In \textbf{Scenario 2}, LOSO results showed that Garmin watches outperformed Empatica E4’s HRV-only model and even its' combined HRV \& EDA model, achieving an AUROC of 0.961, highlighting its effectiveness in capturing physiologial stress. The pre-trained Mishra et al. model further confirmed Garmin Forerunner 55’s robustness, obtaining an AUROC of 0.968, outperforming Empatica E4’s HRV and HRV \& EDA models. Unlike Empatica E4, Garmin is a commercially established consumer wearable with a substantial market share, making it a practical and scalable solution for real-world stress monitoring.

\section{DISCUSSIONS}
\begin{figure}[h] 
    \centering
    \includegraphics[width=\linewidth]{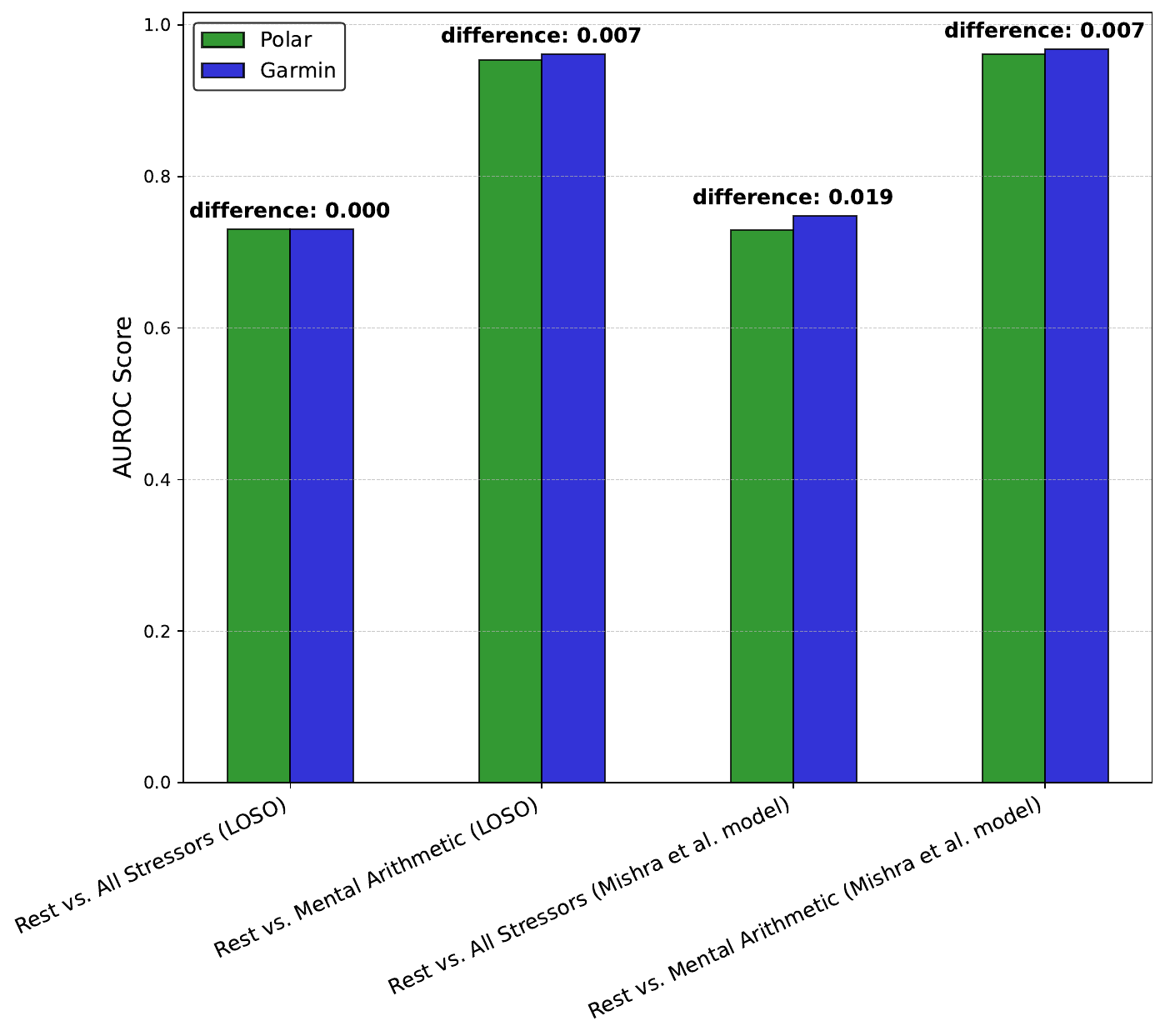} 
    \caption{Stress detection performance comparison of consumer wearable, Garmin Forerunner 55s and the research-grade Polar H10 in a lab setting with minimal differences in AUROC scores. Garmin Forerunner 55s offers greater flexibility for real-world use due to its wrist-worn design, whereas Polar H10's chest strap positioning may limit wearability and convenience for long-term stress monitoring.}
    \label{fig:garmin_polar_compare}
\end{figure}

We observed performance variations across devices and evaluation methods. While our pre-trained model demonstrated strong generalization for certain devices, such as Garmin Forerunner 55s, Polar H10 heart rate monitor, and the gold-standard medical device Biopac MP160, noticeable differences emerged in AUROC scores for Empatica E4, where the model did not generalize as effectively. These variations underscore the importance of device-specific evaluations when assessing consumer-grade stress monitoring solutions. For device-specific reproducibility, as expected, when using both HRV and EDA signals from Biopac MP160, we achieved the best performance, with almost perfect AUROC scores. We observed that data from Garmin devices often led to slightly better performance than the Polar H10 chest monitor illustrated in Fig \ref{fig:garmin_polar_compare}. Both demonstrated high AUROC in the rest vs. mental arithmetic stress task, almost comparable to the Biopac MP160.

The HRV data from Empatica E4 alone had noticeably poorer performance compared to Garmin and Polar devices. However, adding EDA data, led to a significant performance improvement in the LOSO evaluations. While the Mishra~et~al.\ models for HRV+EDA were trained using Empatica E4 data from the previous study, the model does not appear to generalize well with our Empatica E4 data (although the model works well with the Biopac MP160 data). The HRV-only model from Mishra et al., when applied to Garmin Forerunner 55s and the Polar H10 chest monitor, mostly outperforms the HRV+EDA model using Empatica E4. During detecting rest vs. all stressors in LOSO cross-validation, we found that HRV+EDA model performance from Empatica E4 was better than Garmin Forerunner 55s and Polar H10's HRV-only models. In real-world settings, Empatica E4 has been reported to suffer from frequent artifacts and significant data loss in EDA recordings, often caused by connection loss of the electrodes to the skin because of body motions \cite{bottcher2022data}. These limitations reduce its reliability as a data source and make it less suitable for comparison with Garmin Forerunner 55s. 

In our analysis, given that the pre-trained Mishra~et~al. model performs well, with the Biopac MP160 data, we hypothesize that the data quality from Empatica E4 devices could vary significantly with deployments, which could result in the lack of reproducibility of the pre-trained models, but could obtain robust performance in our LOSO evaluations. It's worth mentioning that no additional actions were taken on the Empatica E4 data processing, supporting our hypothesis that its quality might vary significantly across deployments. Additionally, it is interesting to note that the performance from Garmin devices using HRV alone, was on par with Empatica E4 using HRV and EDA data in most evaluations, particularly for detecting the rest vs. mental arithmetic tasks, thus highlighting its effectiveness in stress detection. Garmin watches' superior performance may stem from devices' consistent data validity and reliability, especially for HRV, which relies on accurate heart rate measurements. Garmin Forerunner 55s are slightly more expensive than the Polar H10 chest monitors, however, the lightweight, wrist-worn design of Garmin watches enhances long-term usability and participant comfort compared to the chest-worn Polar, which may be crucial for real-world, continuous stress monitoring.  These results support the feasibility of using HRV data alone from Garmin devices for stress detection. With Garmin ranking fifth in global shipments of activity trackers \cite{evenson2020review}, its consumer-friendly design and widespread adoption highlight its potential for everyday stress monitoring applications.

While our findings provide valuable insights into device-specific performance in stress detection, several limitations should be acknowledged. First, our participant sample consisted of 35 undergraduate students from Northeastern University, which may limit the generalizability of the results to broader populations such as working professionals, older adults, patients, or children. Stress perception, physiological responses, and wearability preferences may differ significantly across these groups. However, in this study, we intentionally focused on undergraduate students, as they are particularly vulnerable to mental health challenges stemming from sustained academic pressure and societal expectations. Second, although we evaluated a diverse set of devices—including a gold-standard medical-grade system, a chest-worn heart rate monitor, and two wrist-worn consumer wearables—we did not include other widely adopted devices such as the Apple Watch, Fitbit, Samsung Galaxy Watch, Huawei, or Xiaomi wearables. Future research should incorporate these high-market-share consumer-grade devices to increase the societal relevance and real-world applicability of stress detection tools. Expanding the participant pool to include more diverse populations will improve the inclusivity and generalizability of wearable-based monitoring. Lastly, larger samples and additional analyses are necessary to assess statistical differences in model performance across individual devices.

\section{CONCLUSIONS}
This paper extends stress detection reproducibility to consumer wearable sensors by evaluating stress detection models beyond traditional research-grade devices. While prior work focused on research devices like Polar H7/H10 and Empatica E4, this study assesses Garmin Forerunner 55, a consumer wearable with a large market share, to determine its potential for stress monitoring. While our prior model demonstrated relatively consistent performance across most devices during in-lab stress detection, it did not generalize well to Empatica E4, highlighting device-specific variations in stress detection. Key findings reveal that while combining HRV and EDA from the Biopac MP160 device yielded the highest stress predictions in lab settings, Garmin wearables demonstrated superior potential in stress monitoring. Empatica E4 performs well when combining HRV and EDA, however, its HRV-only or HRV+EDA models' performance does not surpass Garmin Forerunner 55s, and its data quality might significantly vary across devices with deployments. While both Polar chest-worn and Garmin wrist-worn devices performed comparably, Polar’s chest-worn design poses challenges for long-term stress monitoring due to discomfort and limited wearability. Finally, Garmin Forerunner 55s demonstrated potential with improved stress detection performance, comparable to research-grade devices, with the added advantage of consumer-friendly wearability for continuous monitoring. This research advances digital phenotyping for stress by demonstrating the potential of consumer wearable Garmin Forerunner 55s in improving stress detection reproducibility. Future work should explore integrating wearable and smartphone data to enhance stress prediction models and improve generalizability ``in-the-wild'' settings.

\bibliographystyle{IEEEtran}
\bibliography{root}

\end{document}